\documentclass[useAMS,usenatbib, onecolumn]{mn2e}
\usepackage[utf8]{inputenc}
\usepackage[fleqn]{amsmath}
\usepackage{mathtools}
\usepackage{amssymb}
\usepackage[dvipdfm]{graphicx}
\usepackage{natbib}
\usepackage{float}
\usepackage{gensymb}

\title[Circular polarisation of synchrotron radiation in high magnetic fields]{Circular polarisation of synchrotron radiation in high magnetic fields}
\author[D. de B\'urca \& A. Shearer]{D. de B\'urca$^{1}$\thanks{E-mail: diarmaiddeburca@gmail.com} and A. Shearer$^{1}$\thanks{Email: andy.shearer@nuigalway.ie}\\
$^{1}$Physics Dept., National University of Ireland, Galway, Galway, Ireland}

\begin{document}

\date{Accepted xxxx. Received xxxx; in original form xxxx}

\pagerange{\pageref{firstpage}--\pageref{lastpage}} \pubyear{2015}

\maketitle

2
\label{firstpage}

\begin{abstract}
The general model for incoherent synchrotron radiation has long been known, with the first theory being published by Westfold in $1959$ and continued by Westfold and Legg in $1968$. 
When this model was first developed it was applied to radiation from Jupiter, with a magnetic field of $\approx$ 1 G. Pulsars have a magnetic field of $\approx 10^{12}$ G. The Westfold and 
Legg model predict a circular polarization which is proportional to the square root of the magnetic field, and consequently predicts greater than 100 per cent circular polarization 
at high magnetic fields. Here a new model is derived based upon a more detailed analysis of the pitch angle distribution.  This model is concerned with the frequency range 
$f_{B_0}/\gamma <<f\lesssim f_{B_0}$, noting that $f_{B_0} = 2.7\times10^7B$, which for a relatively high magnetic field ($\sim 10^6-10^8$ Gauss) leaves emission in the optical range. This is 
much lower than the expected frequency peak for a mono-energetic particle of $0.29\frac{3eB}{4\pi m_e c}\gamma^2$.  We predict the circular polarization peaks around $10^7$G in the optical 
regime with the radiation almost $15$ per cent circularly polarized. The linear polarization changes from 
about $60$ to $80$ per cent in the same regime. We examine implications of this for pulsar studies.
\end{abstract}


\section{Introduction}

Pulsar emission in the optical regime is generally accepted to be 
incoherent synchrotron radiation and consequently it should be polarized.  To date most attention has been on linear polarization, in part 
due to instrumental limitations of most polarimeters \cite*{Agaetal2009,Smith1988}. Optical instrumentation, such as the Galway Astronomical 
Stokes Polarimeter (GASP) \cite{Gillian2013}, are now in a position to measure all of the Stokes parameters from pulsars on time-scales 
from milliseconds to hours.  Hence the requirement for a fully self-consistent model for synchrotron radiation in a high magnetic fields.
  
The original model for synchrotron emission was published in $1959$ \cite{Westfold1959}, with other authors coming
to the same conclusions \cite*{LeRoux1961,SazonovGinsburg1968,GinzbergSyr1969}, albeit generally with 
slightly different derivation methods.   This model was then further developed in $1968$ \cite{LeggWestfold1968}, and 
corrections to the model were applied in $1974$ \cite*{GleesonLeggWestfold1974}.   These corrections do not significantly 
alter the circular polarization calculations.   An error in the derivation used was found $1986$ \cite{Singal1986}, but 
this did not change the model predictions.

The motivation behind the Westfold and Legg (hereafter WL) model of synchrotron radiation was to study 
the emission from Jupiter, with a magnetic field of approximately $1$ G.   As such, the behaviour of the 
model was never tested at high magnetic fields.  As pulsars have extremely high magnetic fields ($\sim 10^{12}$G), it 
is important to test the model in high magnetic fields before applying it to pulsar emission.    When the incoherent synchrotron 
emission is calculated at high magnetic fields ($\sim 10^6$ G), the WL formulation predicts a circular polarization greater than $100$ per cent.   
As this is in clear contradiction of reality, a new model for the incoherent synchrotron emission is required.   The most likely reason was in 
the expansion of the particle pitch angle distribution.   In this paper, this is expanded to the next order of magnitude.   

The second section of this paper states the predictions of the previous model \cite{LeggWestfold1968}, and gives the problems inherent in that model.   The third 
section goes through the expansion of the particle pitch angle distribution and the differences that this makes to the intensity, linear polarization and circular polarization.
Finally, the overall effects of each of the different parameters on the intensity are investigated, and some of the possible implications are discussed, particularly in 
relation to pulsar emission.

\section[THE WL model of synchrotron emission]{THE WL MODEL OF SYNCHROTRON EMISSION}
The WL model that is of interest here gives the Stokes parameters for a power-law distribution of electrons with
\begin{align}
N(E) = E^{-p} \hspace{1in} E_1<E<E_2 
\end{align}
 and $N(E)=0$ otherwise as
 \begin{align}
  I =& \frac{k\mu e^2 c}{2\sqrt{2}}\left(\frac{3}{2}\right)^\frac{p}{2}\phi(\theta)\left(f_{B_0}\sin\theta\right)^
  {\frac{p+1}{2}}f^{-\frac{p-1}{2}}\mathcal{J}_{\frac{p+1}{2}} \\
  Q =& \frac{k\mu e^2 c}{2\sqrt{2}}\left(\frac{3}{2}\right)^\frac{p}{2}\phi(\theta)\left(f_{B_0}\sin\theta\right)
  ^{\frac{p+1}{2}}f^{-\frac{p-1}{2}}\mathcal{L}_{\frac{p+1}{2}} \\
  U =& 0 \\
  V =& \frac{k \mu e^2 c}{\sqrt{3}}\left(\frac{3}{2}\right)^\frac{p}{2}\phi\left(\theta\right)
\cot\theta\left(f_{B_0}\sin\theta\right)^{\frac{p}{2}+1}f^{-\frac{p}{2}} \left[\mathcal{R}_{\frac{p}{2}+1}+\left(1+g(\theta)\right)\left(\mathcal{L}_\frac{p}{2} -\frac{1}{2}\mathcal
   {J}_\frac{p}{2}\right)\right]
 \end{align}
 where $f$ is the frequency, $\mu$ is 
 the permeability constant, $p$ is the power-law index, $e$ is the charge of the particle, $\theta$ is the particle pitch angle, $f_{B_0} = eB (2 \pi m c)^{-1}$ is the  fundamental gyro-frequency, 
 $\Phi(\theta)$ is the pitch angle distribution function, 
 and 
 \begin{align} \begin{matrix*}[l]
  \mathcal{J}_\mathrm{n} = \int\limits_0^\infty x^{\mathrm{n}-1}\int\limits_x^\infty K_\frac{5}{3}\left(\nu\right)d\nu dx = \frac{\frac{2}{3}+\mathrm{n}}{\mathrm{n}}\mathcal{L}_\mathrm{n},  &\mathrm{n}>\frac{2}{3}\\
  \mathcal{L}_\mathrm{n} = \int\limits_0^\infty x^{\mathrm{n}-1}K_\frac{2}{3}(x)dx = 2^{\mathrm{n}-2}\Gamma\left(\frac{1}{2}\mathrm{n}-\frac{1}{3}\right)\Gamma
  \left(\frac{1}{2}\mathrm{n}+\frac{1}{3}\right), &\mathrm{n}>\frac{2}{3}\\
  \mathcal{R}_\mathrm{n} = \int\limits_0^\infty x^{\mathrm{n}-1}K_\frac{1}{3}(x)dx = 2^{\mathrm{n}-2}\Gamma\left(\frac{1}{2}\mathrm{n} - \frac{1}{6}\right)\Gamma
  \left(\frac{1}{2}\mathrm{n}+\frac{1}{6}\right),&\mathrm{n}>\frac{1}{3}
   \end{matrix*}  
 \end{align}

\begin{figure}
\centering
  \includegraphics[width=0.75\linewidth]{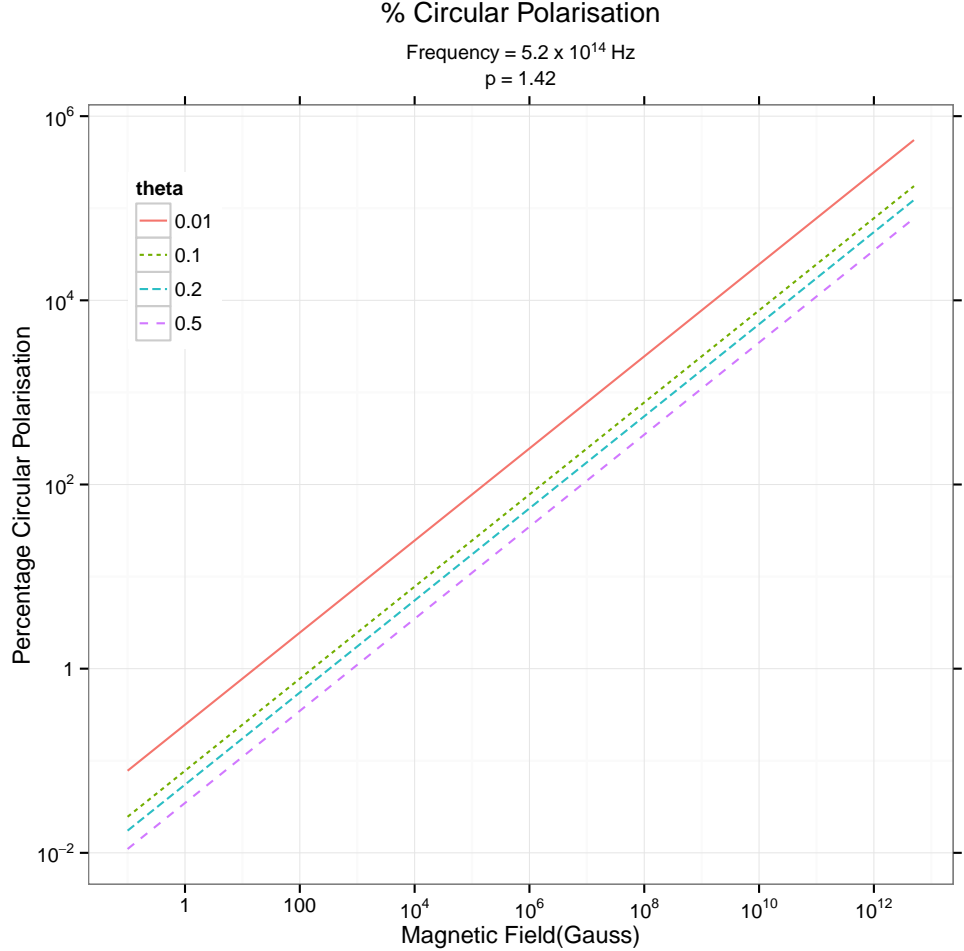}
  \caption{The WL model predicts that the percentage circular polarization will increase linearly with the magnetic field, regardless of the power-law index of the electrons used.
  At some point this model fails and predicts clearly non-physical results.  This is dependant on the pitch angle, the frequency, and the power-law index, but in the area of interest
  the WL model fails long before the predicted surface magnetic field of a pulsar ($\approx 10^{12}$ Gauss). }
  \label{Percentcircpol}
\end{figure}

The circular polarization (defined as $VI^{-1}$) is then given by
\begin{align}
 \frac{V}{I} =& \frac{2\sqrt{2}}{3}\cot\theta\left(f_{B_0}\sin\theta\right)^\frac{1}{2}f^{-\frac{1}{2}}\left[\frac{ \mathcal{R} _{\frac{p}{2}+1}+\left(1+g(\theta)\right) 
 \left(\mathcal{L}_\frac{p}{2} -\frac{1}{2}\mathcal{J}_\frac{p}
 {2}\right)}{\mathcal{J}_{\frac{p+1}{2}}}\right] \propto B^\frac{1}{2}
\end{align}
As the circular polarization is proportional to the root of the magnetic field, it is clear that at some point the degree of circular polarization will exceed $1$.  
This is clearly unrealistic.   However, all models will only have a certain range of validity, and if the magnetic field at which the circular polarization occurs at
extremely high magneitic fields, then the model can still be used for smaller magnetic fields.   The polarization was found to be greater than $100$ per cent at 
approximately $10^5$ to $10^7$ gauss (Fig. \ref{Percentcircpol}) and above, below the surface magnetic field strength of pulsars, but above planetary magnetic fields.   

\section{Expanding the WL Model}
As the model fails at high magnetic fields, a new model is needed to describe the polarization of synchrotron radiation in those fields.   When the derivation used in the WL model is looked at, the main assumptions are 1) that the 
velocity of the particle is close to the speed of light
2) the expansion of the velocity in the frame chosen
3) for a power-law index, the expansion of the distribution of electrons.

In this work we will examine the results when the distribution of electrons is expanded to a higher order.

\subsection{Electric Field}
As synchrotron radiation comes from a source moving in a cyclic fashion, emission will consist of harmonics of the fundamental 
gyro-frequency $f_{B_0}$.   Following the same formulation as WL, the emission from each harmonic can be shown to be 
\begin{align}
 E_\mathrm{n} =& \frac{\mu ce\left(\frac{\omega_B}{b}\right)}{8\pi^2r}\int\limits_0^{\left(\frac{\omega_B}{b}\right)/2\pi}\left[\frac{n\times\left
 (\left(n\times\beta\right)\times\frac{d\beta}{dt}\right)}{\left(1-n\cdot\beta\right)^3}\right] \exp\left[in\left(\frac{\omega_B}{b}\right
 ) t\right]dt \label{E_Nnonretarded}
\end{align}
where the expression in brackets is evaluated at the retarded time 
\begin{align}
 t' = t - \frac{R(t')}{c} \hspace{1in} \mathbf{R}(t') = \mathbf{r}-\mathbf{r}(t') \approx |\mathbf{r}| - \mathbf{n}\cdot\mathbf{r}
\end{align}
{\bf where $b= \beta'\sin\alpha\sin(\alpha-\theta)$ \cite*{LeRoux1961,LeggWestfold1968}. }
Changing the integration to an integration over $t'$, and simplifying, gives
\begin{align}
  E_\mathrm{n} =& \frac{\mu ce}{8\pi^2r}\left(\frac{\omega_B}{b}\right)^2 i n \exp\left(in\left(\frac{\omega_B}{b}\right) \frac{|r|}{c} \right)
 \label{ElectricFieldOne}\int\limits^{\infty}_{-\infty}\mathbf{n}\times\left(\mathbf{n}\times\beta\right)
  \exp\left[in\left(\frac{\omega_B}{b}\right)\left(t'-\frac{\mathbf{n}\cdot\mathbf{r}_0(t')}{c}\right)\right]dt' 
\end{align}

\subsection{Coordinate System}
In order to solve equation (\ref{ElectricFieldOne}), it is necessary to choose a system of coordinates.   In this case, the system will be constructed as follows:  the particle is spiralling around a magnetic field at an angular 
frequency of $\omega_B = qB(\gamma m c)^{-1}$, where $q$ is the charge, $B$ is the strength of the magnetic field and $\gamma$ is the Lorentz
factor of the particle.   The particle maintains a constant pitch angle of $\alpha$ with respect to the the magnetic 
field direction.   At any particular time the orbit has a radius of curvature of \textbf{\textit{a}}.   Now, let the \textbf{\textit{x}}-\textbf{\textit{y}} plane be the instantaneous
plane of the orbit of the particle.   Now, take the origin of the \textbf{\textit{x}}-axis to be the point where the velocity vector and the observer are in 
the \textbf{\textit{x}}-\textbf{\textit{z}} plane, and let the \textbf{\textit{y}} coordinate be in the direction of the radial vector \textbf{\textit{a}}, with the \textbf{\textit{x}} coordinate being defined as perpendicular
to the \textit{\textbf{y}} and \textbf{\textit{z}} coordinates.
\begin{figure}
\centering
 \includegraphics[width=0.5\linewidth]{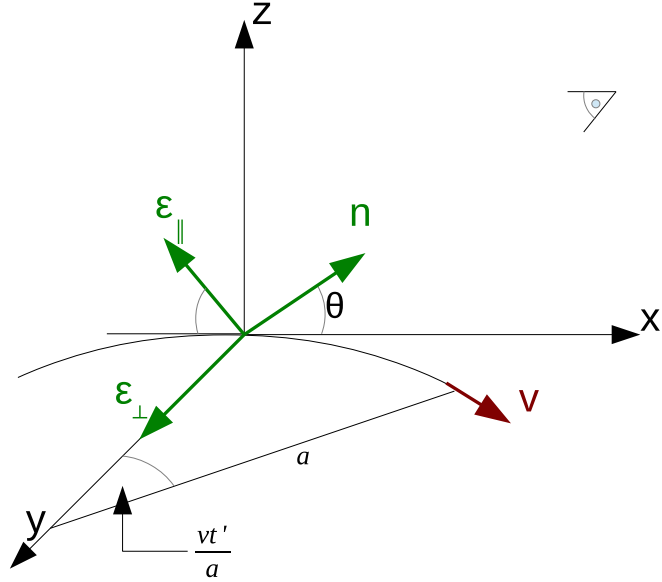}
 \caption{The geometry used in order to calculate the synchrotron emission.  Firstly, define the \textbf{\textit{x}}-\textbf{\textit{y}} plane as the instantaneous plane of orbit of the particle.  Then define the origin as the point at which 
 the velocity \textit{\textbf{v}} and the vector to the observer \textit{\textbf{n}} are both in the \textbf{\textit{x}}-\textbf{\textit{z}} plane.  Define $\boldsymbol{\varepsilon}_\perp$ to be along the y axis, and 
 $\boldsymbol{\varepsilon}_\|$ as \textbf{\textit{n}}$\boldsymbol\times\boldsymbol{\varepsilon}_\perp$. This gives a natural frame of reference for the polarization of the emission.
}
 \label{Coordsys}
\end{figure}

Now, define a new set of coordinates (\textbf{\textit{n}}, $\boldsymbol{\varepsilon}_\|,\boldsymbol{\varepsilon}_\perp$) such that the origin is at the same point
as the (\textbf{\textit{x}},\textbf{\textit{y}},\textbf{\textit{z}}) coordinate system origin, \textbf{\textit{n}} is pointing towards the observer, $\boldsymbol{\varepsilon}_\perp$ is pointing along \textbf{\textit{y}}, and 
$\boldsymbol{\varepsilon}_\| = $\textbf{\textit{n}}$\boldsymbol\times\boldsymbol{\varepsilon}_\perp$.   This then gives a natural coordinate system in which to consider the
polarization of the emission, as $\boldsymbol{\varepsilon}_\perp$ is perpendicular to the magnetic field and $\boldsymbol{\varepsilon}_\|$ is parallel to 
the magnetic field direction, as seen in projection by an observer, as can be seen in Fig. \ref{Coordsys}.  {\bf Here $a$ stands for the radius of curvature of the particle.} 

Finally, in this coordinate system the velocity and acceleration are \cite{Longair}
\begin{align}
 \boldsymbol{\mathit{r}}_0(t') =& 2a\sin{\left(\frac{vt'}{2a}\right)}\Bigg[\sin{\left(\frac{vt'}{2a}\right)}\boldsymbol{\varepsilon}_\perp+ \cos\theta\cos\left( \frac{vt'}{2a}\right) \boldsymbol{\mathit{{n}}} - \sin\theta\cos\left(\frac{vt'}{2a}\right)\boldsymbol{\varepsilon}_\| \Bigg] \\
 \boldsymbol{\mathit{v}} =& v\Bigg[\sin{\left(\frac{vt'}{a}\right)}\boldsymbol{\varepsilon}_\perp + \cos{\theta}\cos{\left(\frac{vt'}{a}\right)} \boldsymbol{\mathit{n}}
 - \sin{\theta}\cos{\left(\frac{vt'}{a}\right)}\boldsymbol{\varepsilon}_\|\Bigg]
\end{align}

This then gives the electric field (splitting it into its component parts parallel and perpendicular to the projection of the magnetic field), and dropping the subscript, 
\begin{align}
  E_\| =&  \frac{\mu ce}{8\pi^2r}\left(\frac{\omega_B}{b}\right)^2 i n \exp\left[in\left(\frac{\omega_B}{b}\right]\frac{|r|}{c}\right)\int
\limits_{-\infty}^\infty \theta \exp\left[H\right]dt' \\
 E_\perp =& -\frac{\mu ce}{8\pi^2r}\left(\frac{\omega_B}{b}\right)^2 i n \exp\left[in\left(\frac{\omega_B}{b}\right]\frac{|r|}{c}\right) \int
 \limits_{-\infty}^\infty\frac{vt'}{a}\exp\left[\mathcal{H}\right]dt' \\
 \mathcal{H}=&in\left(\frac{\omega_B}{b}\right)\left[t'\left(1-\frac{v}{c}\right) +\frac{v}{c}\frac{\theta^2}{2}t' +\frac{v^3}{6ca^2}t'^3 \right]
\end{align}
A convenient substitution of
\begin{align}
 \theta_\gamma^2 = (1-\gamma^2\theta^2); \hspace{1in} y = \frac{\gamma ct'}{a\theta_\gamma}; \hspace{1in} \eta = \frac{n\left(
 \frac{\omega_B}{b}\right)a \theta_\gamma^3}{3c\gamma^3}
\end{align}
then gives

\begin{align}
 E_\| =&  \frac{\mu ce}{8\pi^2r}\left(\frac{\omega_B}{b}\right)^2 i n \exp\left[in\left(\frac{\omega_B}{b}\right)\frac{|r|}{c}\right]
 \left(\frac{a\theta\theta_\gamma}{\gamma c}\right)\int\limits_{-\infty}^\infty \exp\left[i\eta\left(y+\frac{1}{3}y^3\right)\right]dy \\
 E_\perp =& -\frac{\mu ce}{8\pi^2r}\left(\frac{\omega_B}{b}\right)^2 i n \exp\left[in\left(\frac{\omega_B}{b}\right)\frac{|r|}{c}\right]\left(\frac{a\theta_\gamma}{\gamma c}\right)^2\int\limits_{-\infty}^\infty
 y \exp\left[i\eta\left(y+\frac{1}{3}y^3\right)\right]dy
\end{align}

\subsection{Emission-polarization tensor}
The emission polarization tensor is defined as
\begin{align}
 \rho = \frac{2\pi r^2}{\mu}\left(\begin{array}{cc}
E_\perp E_\perp^* & E_\perp E_\|^* \\
E_\| E_\perp^* & E_\| E_\|^*
\end{array}\right)\label{PolarisationTensor}
\end{align}
This is equivalent to getting the Stokes parameters for each harmonic, as
\begin{align}
 I &= \rho_{11}+\rho_{22} \\
Q &= \rho_{11}-\rho_{22} \\
U &= \rho_{12}+\rho_{21} \\
V &= \frac{1}{i}\left(\rho_{12}-\rho_{21}\right)
\end{align}

\subsection{Airy Functions}
It is possible to convert the electric field exponential into Bessel functions (see supplementary materials).   This gives
\begin{align}
 \rho_{11} =& \frac{\mu e^2 c}{24\pi^4} \left(\frac{\omega_B}{b}\right)^4n^2 \frac{a^2}{c^2}\frac{\theta_\gamma^4}{\gamma^4}K^2_\frac{2}{3}\left(\eta\right)\\
 \rho_{12}=& \frac{\mu e^2 c}{24\pi^4} \left(\frac{\omega_B}{b}\right)^4n^2 \frac{\theta\theta_\gamma}{\gamma^3}\frac{a^2}{c^2}K_\frac{1}{3}\left(\eta\right)K_\frac{2}{3}\left(\eta\right)\\
 \rho_{22} =& \frac{\mu e^2 c}{24\pi^4} \left(\frac{\omega_B}{b}\right)^4n^2\theta^2\frac{a^2}{c^2}K^2_\frac{1}{3}\left(\eta\right) 
\end{align}
This then gives the polarization tensor for a particular harmonic of the emission.   

\subsection{Converting to Frequency Domain}
For large-order harmonics, the radiation becomes quasi-continuous \cite{LeggWestfold1968} and it is possible to convert 
the polarization tensor for a single harmonic to the frequency polarization tensor using
\begin{align}
 \rho_f = \frac{\rho_\mathrm{n} b}{f_B} \label{harmtofreq1}
\end{align}
where $f_B = f_{B_0}/\gamma$, $b=\beta'\sin\alpha \sin\left(\alpha - \theta\right)$, $\rho_\mathrm{n}$ is the polarization tensor for a single harmonic and $\rho_f$ is the polarization emission tensor at a particular frequency,
and 
\begin{align}
 f = nf_B b^{-1}\label{harmtofreq2}
\end{align}

This gives the polarization tensor at a particular frequency.   It is convenient to convert from the frequency into a dimensionless parameter $x$ such that

 \begin{align}
 x= \frac{f}{f_\mathrm{c}} = \frac{4\pi a}{3c\gamma^3}f \label{xintermsoff}
\end{align}
which in turn gives
\begin{align}
 n = \frac{3c\gamma^3}{2a}\left(\frac{b}{\omega_B}\right)x \hspace{1in} \eta=\frac{x}{2}\theta_\gamma^3
\end{align}

This gives
\begin{align}
  \rho_{x_{11}} =&\frac{3}{16}\frac{\mu e^2 c}{\pi^3}\frac{\omega_B}{b}x^2 \gamma^2\theta_\gamma^4K_\frac{2}{3}^2\left(\frac{x}{2}\theta_\gamma^3\right) \\
 \rho_{x_{12}} =& \frac{3}{16}\frac{\mu e^2 c}{\pi^3}\frac{\omega_B}{b}x^2 \gamma^3\theta\theta_\gamma^3K_\frac{1}{3}\left(\frac{x}{2}\theta_\gamma^3\right)K_\frac{2}{3}\left(\frac{x}{2}\theta_\gamma^3\right)\\
 \rho_{x_{22}} =& \frac{3}{16}\frac{\mu e^2 c}{\pi^3}\frac{\omega_B}{b}x^2\gamma^4\theta_\gamma^2\theta^2K_\frac{1}{3}^2\left(\frac{x}{2}\theta_\gamma^3\right)
\end{align}

\subsection{Power-law Polarisation-emission tensor}
When there is a power-law of particles, the polarization-emission tensor for that population of particles is
\begin{align}
  n_x(n) = 2\pi\int\limits_0^\infty N(E)\int\limits^{\Omega(n)} \phi(\alpha)\sin\alpha b P_x(n)d\Omega(n) dE \label{Polemissiontensor}
\end{align}
In order to solve this it is possible to represent $\alpha$ as $\alpha +\theta$. Then the solid angle is represented as $d\Omega(\mathbf{n})$ as $2\pi\sin\alpha d\theta$.   
This can be written as  $\alpha = \alpha'+ \theta$ and substituted into equation (\ref{Polemissiontensor}).   To third order, the particle pitch angle distribution can be written as
\begin{align}
 \phi(\alpha'+\theta)\sin\left(\alpha'+\theta\right) = f(\alpha') +g(\alpha')\theta+h(\alpha')\theta^2
\end{align}
where
\begin{align}
 f(\alpha') =& \phi(\alpha')\sin\alpha' \\
 g(\alpha') =& \phi'(\alpha')\sin\alpha' +\phi(\alpha')\cos\alpha'\\
 h(\alpha)  =& \phi''(\alpha')\sin\alpha' +2\phi'(\alpha')\cos\alpha'-\phi(\alpha')\sin\alpha'
\end{align}
and $\phi(\alpha)$ is the pitch angle distribution of the particles itself.   This gives the polarization-emission tensor as (writing $\alpha'$ as $\alpha$, see supplementary materials) 
\begin{align}
 n_{x_{11}} =&\frac{1}{4\sqrt{2}}\mu e^2 c f_{B_{0}}^\frac{p+1}{2} \left(\frac{3}{2}\right)^\frac{p}{2}\sin\theta^\frac{p+1}{2}f^\frac{1-p}{2}\Bigg[\phi(\alpha)\left(\mathcal{J}_\frac{p+1}{2} +\mathcal{L}_\frac{p+1}{2}\right)  + 
 \frac{3h(\alpha)}{2}\left(\frac{f_{B_0}}{f}\right)\left(3\mathcal{Q}_\frac{p+3}{2} -2\mathcal{L}_\frac{p+3}{2} -\mathcal{J}_\frac{p+3}{2}\right) \Bigg]\\
 n_{x_{12}} =& \frac{1}{4\sqrt{2}}{\mu e^2 c} g(\alpha)\left(\frac{3}{2}\right)^\frac{p}{2}f_{B_0}^{\frac{p}{2}+1}f^{\frac{-p}{2}}\sin\theta^{\frac{p}{2}}\left(2\mathcal{L}_{\frac{p}{2}+1}-\mathcal{J}_{\frac{p}{2}+1}\right)\\
  n_{x_{22}} =&\frac{1}{4\sqrt{2}}\mu e^2 c f_{B_{0}}^\frac{p+1}{2} \left(\frac{3}{2}\right)^\frac{p}{2}\sin\theta^\frac{p+1}{2}f^\frac{1-p}{2}\Bigg[\phi(\alpha)\left(\mathcal{J}_\frac{p+1}{2} -\mathcal{L}_\frac{p+1}{2}\right)  
 +\frac{9h(\alpha)}{8}\left(\mathcal{Q}_\frac{p+3}{2} -\mathcal{J}_\frac{p+3}{2}\right)\Bigg]
\end{align}
where 
\begin{align}
 \mathcal{Q}_n = \int\limits_0^\infty x^{n-1}K_\frac{4}{3}\left(x\right)dx
\end{align}
and the other parameters are as before.

\section{Discussion}
\begin{figure}
\includegraphics[width=\linewidth]{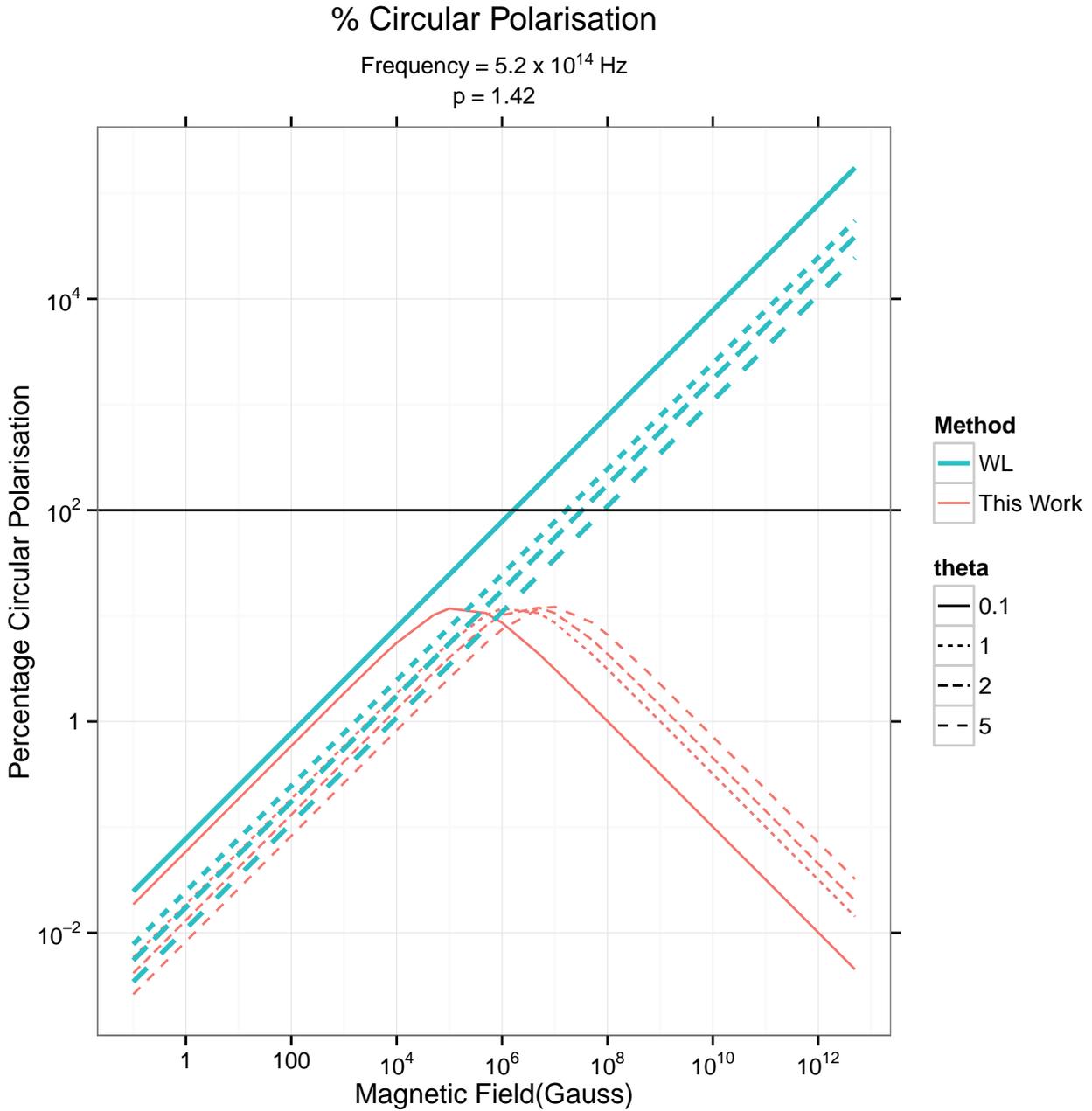}
\caption{The circular polarization for a power-law distribution of particles with a power law index of $1.42$, at a frequency of $5.212 \times10^{14}$ Hz.   Here WL stands for Westfold and Legg, the original emission theory, 
and theta stands for the particle pitch angle.}
\label{Circpolpic}
\end{figure}

\begin{figure}
 \includegraphics[width=\linewidth]{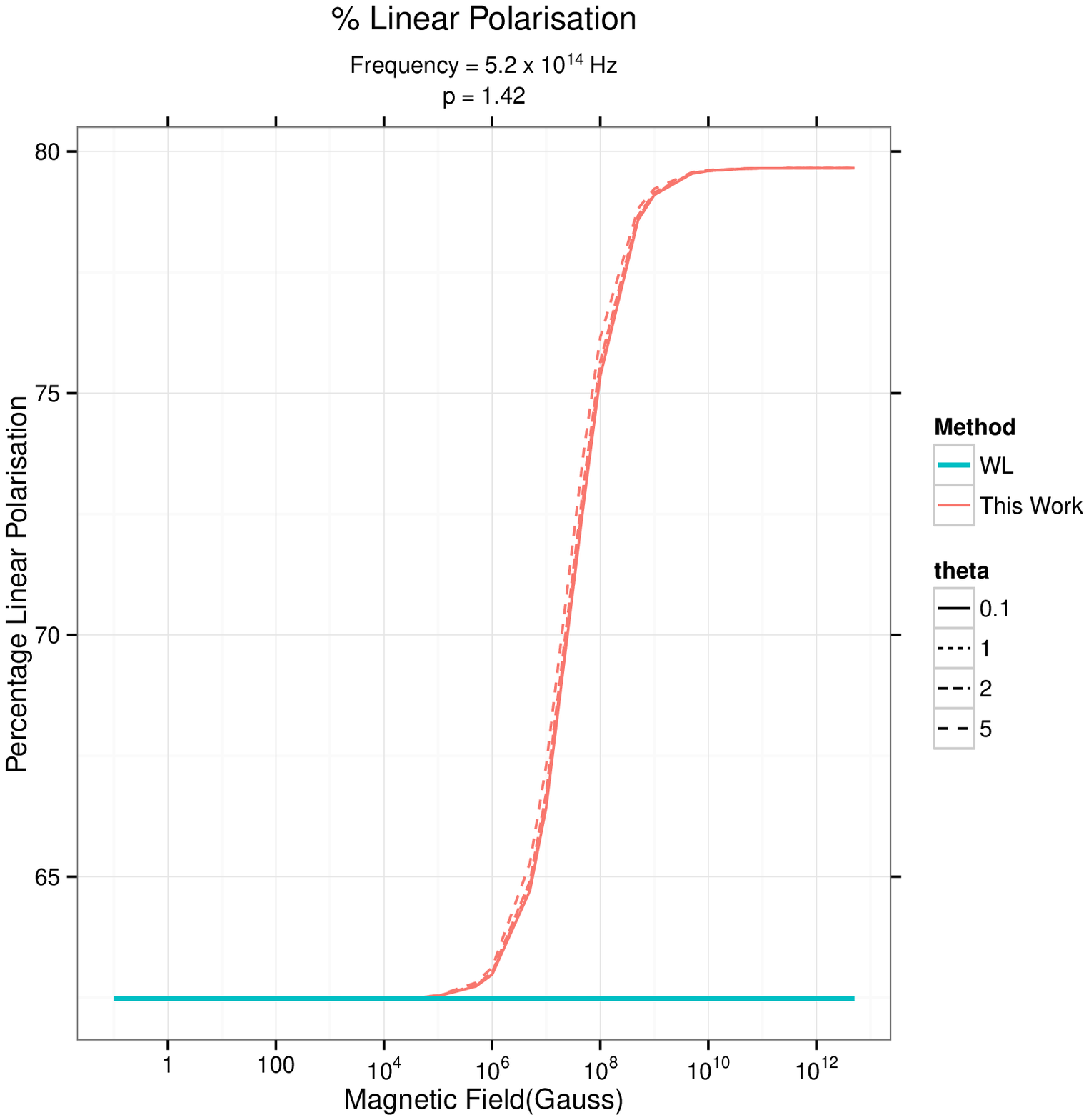}
 \caption{The linear polarization change with regard to the magnetic field for a particle power law index of $1.42$ and frequency $5.212\times10^{14}$ Hz and theta for the particle pitch angle..   As can be seen, the 
 linear polarization is steady at low magnetic  fields and at high magnetic fields, with the linear polarization changing smoothly between the two values in intermediate range of magnetic field values.    }
 \label{Linpolpic}
\end{figure}

This formulation predicts that the cirular polarisation will not exceed one hundred per cent.   For a particle power law index of $1.42$, the circular polarization remains less than fifteen percent for 
all magnetic field values in the optical regime (Fig. \ref{Circpolpic}), 
while the linear polarization changes value at high magnetic fields (Fig. \ref{Linpolpic})
but remains at a steady value except in the intermediate range of magnetic fields ($10^6- 10^{10}$ G).   One result is that the level of circular polarization is directly dependant on the magnetic field.   
As such, measurements of the circular polarization gives some potential constraints to the magnetic field strength of 
the emission volume.  

There is one major constraint to this emission model.   In order to obtain emission at high magnetic fields, the $h(\alpha)$ term has to be positive.
In general, for small values of $\theta$, this can be obtained by considering particle pitch angle distributions with positive first derivatives.   So, in this paper the particle pitch angle distribution 
used $\phi(\alpha) = \sin(\alpha)sin(\alpha_{max})^{-1}$.   However, this particle pitch angle distribution does not in general agree with the particle pitch angle distributions that are predicted (\textrm{e.g.} guassian).   
Physically, this type of distribution would result from particles which lose more energy the
closer they are to the magnetic field line.

Another area of interest would be in pulsar studies.   There are currently a number of different theories about pulsar high-energy/optical emission \cite*{Cheng1986, Outergap, Stripedwind, AnnularGap, Harding2013} which agree on the process of emission, pair production 
creating a plasma which then emits using synchrotron radiation, but which disagree on the location of the pulsar emission zone.   As the magnetic field strength can be correlated with the position in the pulsar magnetosphere, this 
provides a test to constrain the pulsar emission location.  One method to constrain the pulsar emission locations would be to use an inverse mapping approach, \cite{POREC}. They considered emission from all 
parts of the magnetosphere, and compared that emission  to optical observations.   They found that the majority of the emission came from approximately $300$ km from the pulsar surface, where the magnetic field strength is in the range of 
$10^{7}-10^{8}$ G.  Future work would involve incorporating our model into the code and checking if the emission areas change significantly.
 
The linear polarization predicted by our method is not dependant on the particle pitch angle except in intermediate magnetic fields, and in certain regimes is not dependant on the magnetic field strength.   However, 
it is very sensitive to the particle power-law index.   As such, it could be possible to constrain the particle power-law index from measurements of the linear polarization. A combined measure of linear and circular 
polarization is therefore an important diagnostic tool for determining the geometry of pulsar emission zones.
  
\begin{figure}
 \includegraphics[width=0.75\linewidth]{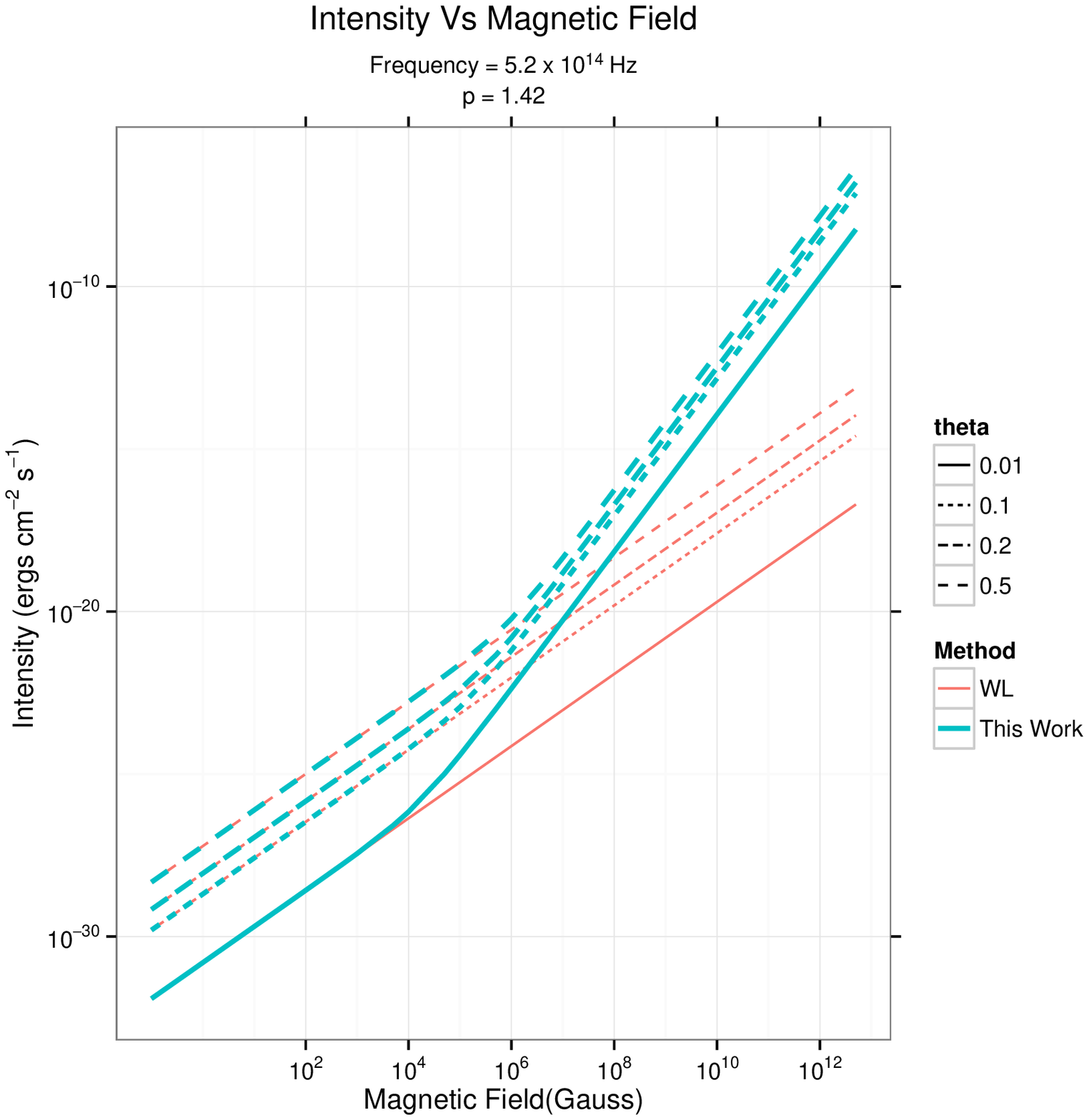}
 \caption{A comparison of the intensities predicted by both models.  The frequency is $5.212\times10^{14}$ Hz, the pitch angle distribution is $\Phi(\alpha) = \sin(\alpha)$, 
 the particle power-law index is 1.42, and theta stands for the pitch angle.   As can be seen, at approximately $10^{4}-10^{6}$ G, the slope of the intensity changes from $\alpha = 0.5(p-1)$ to $0.5(p+1)$.}
 \label{Intensitychange}
\end{figure}

The relationship between the observed power-law index, and the particle power-law index, is different at high magnetic fields.   This relationship has been accepted as $p=2\alpha+1$ \cite*{Longair, RybickiLightman}, however, at high 
magnetic fields, this relationship changes to $p = 2\alpha-1$ in this formulation  (Fig. \ref{Intensitychange}).   

 There are still a number of issues to be addressed. The frequencies of interest here are far from the maximal spectral frequency of a single shape.   Therefore, it is worthwhile to note that there could be
 errors introduced due to the integration going between $0$ and $\infty$ rather than over a realistic energy range.   Investigating the error is beyond the scope of this work.   To first order we 
 can state that the polarisation measurements are correct, as calculating the formula for $\mathcal{L}_n$ and $\mathcal{J}_n$ \cite{GleesonLeggWestfold1974} gives a ratio of $0.644$ for the integration over $0$ to $\infty$ , and a 
 slowly varying ratio between $0.644$-$0.500$ (as the magnetic field is increases from $10^5$-$10^{11}$) for the ratio from the exactly calculated values.  Lifetime effects also limit the effective energy range - 
if the energy is below a minimum of $\gamma = f_{b_0}/(f\sin^2\theta)$ then there can be no radiation \cite{GleesonLeggWestfold1974}.   At high magnetic fields this can be of the order of $\gamma = 10^3$.


Measurements of the linear and circular polarization from pulsars with apparent magnitudes 
less than $25$ is possible with instruments like GASP on $4$-m class telescopes. Our predictions can therefore be tested on normal pulsars such as the Crab pulsar and on magnetars such as 4U0142+61. We also develop our inverse mapping 
approach \cite{POREC} to include circular polarization.  

We note that for the Crab pulsar the maximum linear polarisation is fifteen per cent \cite{Agaetal2009}, whereas we predict higher values ($>60$ per cent), consistent with the WL formulation.   This discrepancy requires further investigation
and could be due to either the impact of different pitch angle distribution or a more astrophysical explanation.   Future work should clarify this.

\section{Acknowledgements}
The authors are grateful to the anonymous referee whose comments improved an earlier version of this paper.   DdeB acknowledges the National University of Galway, Ireland's College of Science PhD scholarship which funded this work.

\bibliographystyle{mn2e_new}

\bsp

\label{lastpage}

\end{document}